\documentclass[showpacs,showkeys,preprintnumbers,amsmath,amssymb,latexsym]{revtex4} 

\usepackage{graphicx}
\usepackage{dcolumn}

\def\rmd{{\rm d}} 
\def\fl{{}}

\def\beq{\begin{equation}}

\catcode`@=11  
\def\meqalign#1{\null\,\vcenter{\openup\jot\m@th
\ialign{\strut\hfil$\displaystyle{##}$&&$\displaystyle{{}##}$\hfil
\crcr#1\crcr}}\,}
\catcode`@=12  

\def\pmb#1{\setbox0=\hbox{$#1$}%
  \kern-.025em\copy0\kern-\wd0
  \kern.05em\copy0\kern-\wd0
  \kern-.025em\raise.0433em\box0}
 
\begin{document}

\title{Inertial effects of an accelerating black hole}






\author{D. Bini}
\affiliation{Istituto per le Applicazioni del Calcolo ``M. Picone'', CNR, I-00161 Rome, Italy and\\ 
International Center for Relativistic Astrophysics - I.C.R.A.\\
University of Rome ``La Sapienza'', I-00185 Rome, Italy}

\author{C. Cherubini}
\affiliation{ Faculty of Engineering, University Campus Bio-Medico of Rome, 
via E. Longoni 47, 00155 Rome, Italy,\\
International Center for Relativistic Astrophysics - I.C.R.A.\\
University of Rome ``La Sapienza'', I-00185 Rome, Italy}

\author{B. Mashhoon}
\affiliation{Department of Physics and Astronomy,
University of Missouri-Columbia, Columbia,
Missouri 65211, USA}

\date{\today}

\begin{abstract}
We consider the static vacuum C metric that represents the 
gravitational field of a black hole of mass $m$ undergoing uniform 
translational acceleration $A$ such that $mA<1/(3\sqrt{3})$. 
The influence of the inertial acceleration on the exterior 
perturbations of this background are investigated. 
In particular, we find no evidence for a direct spin-acceleration coupling.
\end{abstract}



\maketitle

\section{Introduction}

We study the motion of test particles and the 
propagation of wave fields on the exterior vacuum 
C metric background, which can be thought of as a 
nonlinear superposition of Schwarzschild and Rindler spacetimes. 
We find geodesic orbits that are circles 
about the direction of acceleration. Moreover, 
we consider the massless field perturbations of 
the C metric in search of a {\it direct coupling} between 
the spin of the perturbing field and the acceleration of the background, 
in complete analogy with the well-known spin-rotation coupling \cite{srcoupl}. 
The results indicate that such a coupling does not exist. 
Furthermore, in the linear approximation to the C metric,
 we show that the propagation of the scalar field on this 
background entails a \lq\lq gravitational Stark effect" that is analogous to the motion of an electron in the Stark effect.

\section{Vacuum C Metric}

The vacuum C metric was first discovered by Levi-Civita \cite{LC} in 1918 within a class of Petrov type D (degenerate) static vacuum metrics.
However, over the years it has been rediscovered many times: by Newman and Tamburino \cite{newtam} in 1961, by Robinson and Trautman \cite{robtra} in 1961 and again by Ehlers and Kundt \cite{ehlkun} ---who called it the C metric---
in 1962. The charged C metric has been studied in detail by Kinnersley and Walker \cite{kin69,kinwal}. In general the spacetime represented by the C metric contains one or, via an extension, two uniformly accelerated particles as explained in \cite{kinwal,bon83}.
A description 
of the geometric properties as well as the various extensions of the C metric is contained in \cite{ES}, which should be consulted for a more complete list of references.
The main property of the C metric is the existence of two hypersurface-orthogonal  Killing vectors, one of which is
timelike (showing the static property of the metric) in the spacetime region of interest in this work.
The most familiar form of the  C metric is  \cite{kin69,kinwal}
\begin{equation}
\label{met_txyz}
\rmd s^2=\frac{-1}{A^2(\tilde x+\tilde  y)^2}[(\tilde F \rmd t^2 - \tilde F^{-1}\rmd \tilde y{}^2) -(\tilde G^{-1} \rmd \tilde x{}^2 + \tilde G\rmd \tilde z{}^2)],
\end{equation}
where
\begin{equation}
\tilde F(\tilde y)= -1+\tilde y{}^2-2mA\tilde y{}^3,\qquad  \tilde G(\tilde  x)= 1-\tilde x{}^2-2mA\tilde x{}^3, \qquad \tilde G(\tilde x)=-\tilde F(-\tilde x).
\end{equation}
These coordinates are adapted to the hypersurface-orthogonal Killing vector $\kappa=\partial_t$, the spacelike Killing vector
$\partial_{\tilde z}$ and $\partial_{\tilde x}$, which is aligned along the non-degenerate eigenvector of the hypersurface Ricci tensor.
The constants $m\ge 0$ and $A\ge 0$ denote the mass and acceleration of the source, respectively. 
Unless specified otherwise, we choose units such that the gravitational constant and the speed of light in vacuum are unity.
Moreover, we assume that the C metric has signature +2; to preserve this signature, we must have $\tilde G>0$. We assume further that $\tilde F>0$; it turns out that the physical region of interest in this case corresponds to $mA<1/(3\sqrt{3})$ \cite{Farh,pavda,podol} .

Working with the metric in the form (\ref{met_txyz}), the Schwarzschild limit ($A=0$) is not immediate. Therefore, it is useful to introduce the retarded time coordinate $u$, the radial coordinate $r$ and the azimuthal coordinate $\phi$:
\begin{equation}
u=\frac1A [t+\int^{\tilde y} \tilde F^{-1}\rmd \tilde y], \qquad r=\frac{1}{A(\tilde  x+\tilde  y)}, \qquad \phi= \tilde z ,
\end{equation}
so that the metric can be cast in the form
\begin{equation}
\label{cmetEF}
\rmd s^2= -\tilde H \rmd u^2 - 2 \rmd u \rmd r - 2A r^2 \rmd u \rmd \tilde x +\frac{r^2}{\tilde G} \rmd \tilde x{}^2 +r^2\tilde G \rmd \phi^2 , 
\end{equation}
where
\begin{equation}
\tilde H(r,\tilde x)=1-\frac{2m}{r}-A^2r^2 (1-\tilde x{}^2-2mA\tilde x{}^3)-Ar(2\tilde x+6mA\tilde x{}^2)+6mA\tilde  x .
\end{equation}
The norm of the hypersurface-orthogonal Killing vector $\kappa$ is determined by $\tilde H$, 
$\kappa_\alpha \kappa^\alpha = -r^2 \tilde F=-\tilde H/ A^2$,
so that this Killing vector is timelike for $\tilde H>0$.
We find it
convenient to work with the $\{u,r,\theta,\phi \}$ coordinate system, where 
$(r,\theta , \phi)$ are spherical polar coordinates with  $\tilde x=\cos \theta$. Thus the C metric takes the form
\begin{equation}
\label{cmetu}
\rmd s^2= -H \rmd u^2 - 2 \rmd u \rmd r + 2A r^2 \sin \theta \rmd u \rmd \theta +\frac{r^2\sin^2\theta }{G} \rmd \theta^2 +r^2 G \rmd \phi^2  ,
\end{equation} 
where $G$ and $H$ are given by
\begin{eqnarray}
G(\theta)&=& \sin^2 \theta -2mA \cos^3\theta\, , \nonumber \\
H(r,\theta)&=&1-\frac{2m}{r}-A^2r^2 (\sin^2\theta-2mA\cos^3\theta)-2Ar\cos\theta(1+3mA\cos\theta)+6mA\cos\theta .
\end{eqnarray}

To study the location of horizons it is useful to introduce an acceleration length scale based on $A>0$ given by
$L_A= 1/(3\sqrt{3}A)$. It turns out that the modification of the horizons is related to the ratio
of $m$ and $L_A$.
The event horizons of the vacuum C metric are Killing horizons given by $H=0$ \cite{kinwal}.
The solution of $H=0$ can be written as 
$r^{-1}=A (\cos \theta + W^{-1})$,
where $W$ is a solution of $W^3-W+2mA=0$. There are three cases depending on whether $m$ is less than, equal to or greater than $L_A$. We have assumed at the outset that $m<L_A$; therefore, we expect that the two horizons of the Schwarzschild ($r=2m$) and the Rindler ($r=[A(1+\cos\theta)]^{-1}$) metrics will be somewhat modified. In fact let
\begin{equation}
\frac{1}{\sqrt{3}}\left( -\frac{m}{L_A}+i\sqrt{1-\frac{m^2}{L_A^2}}\right)^{1/3} =\hat U+i \hat V,
\end{equation}
then there are three real solutions for $W$ given by
$W=2\hat U$, which results in $r=2m$ for $A\to 0$, $W=-\hat U+\sqrt{3}\hat V$, which results in $r^{-1}=A(1+\cos\theta)$ for $m\to 0$,
and  $W=-\hat U-\sqrt{3}\hat V$, which results in $r^{-1}=A(\cos\theta -1 )$ for $m\to 0$ and is therefore unacceptable.
In the next two sections we will discuss the motion of test particles and the propagation of wave fields in the exterior spacetime region.

\section{Test particle motion: circular orbits}

Imagine the exterior of a spherically symmetric gravitational 
source that is uniformly accelerated along the $\theta=\pi$ direction with acceleration $A$. 
In the rest frame of the source, it is possible to find circular 
orbits about the direction of acceleration. In fact, 
in the Newtonian limit, a test particle can follow such an orbit of 
radius $r \sin \theta$ for fixed $r$ and $\theta$ in the natural spherical polar 
coordinate system $(r, \theta, \phi)$. In this case, $(m/r^2) \cos \theta = A$ and 
the speed of the circular motion $v$ is given by $v^2 = (m/r) \sin^2 \theta$. 
It follows that circular orbits are possible for  $0 < \theta < \pi/2$. 
The situation in general relativity is very similar, but somewhat more complicated. 
Indeed, timelike circular orbits exist for $\theta_0 < \theta < \pi/2$, 
where $G (\theta_0) = 0$. Moreover, for $\theta = \pi/2$, the circular orbit 
is null and is given by $r = 3 m$ for all $A$ such that $p\equiv mA<1/(3\sqrt{3})$. 
Finally, there are spacelike circular orbits for $\pi/2 < \theta < \theta_c$, 
where $\theta_c (p)$ is a critical polar angle; for details, see \cite{bcgj}.

\section{Wave motion: Perturbations}

A master equation, analogous to the one derived in the Kerr spacetime \cite{Guven,teuk1, teuk2, Detweiler, Wainwright, RARITA2,Finley2} and describing massless field perturbations of any spin, has been studied by Prestidge \cite{prest} on the C-metric background. However the physical content of this equation  is not yet completely understood,  because  the master equation cannot be integrated exactly but only separated in $\{t,\tilde x,\tilde y,\tilde z \}$ coordinates.

We present the master equation  for the C metric in a slightly different form compared with the one obtained by Prestidge \cite{prest}. In fact, we use here a principal NP frame which is also Kinnersley-like, i. e. it has the  NP spin coefficient $\epsilon=0$. This allows some further simplification and puts this development in a form very close to the black hole case, where  the master equation formalism has been successfully developed. Details for the derivation of the master equation in this case can be found in \cite{bcmprd,bcmcqg}.

With the C metric in the form (\ref{met_txyz}) and switching the signature to $-2$ to agree with the standard Newman-Penrose formalism, a Kinnersley-like NP principal null tetrad can be easily constructed
with 
\begin{eqnarray}
\mathbf{l}= A(\tilde x +\tilde y)^2  \left(\frac{1}{\tilde F} \partial_t + \partial_{\tilde y}\right),\quad
\mathbf{n}=  \frac{A}{2} \left( \partial_t -\tilde F\partial_{\tilde y}\right ) ,\quad
\mathbf{m}=\frac{\tilde G^{1/2}A(\tilde x+ \tilde y)}{\sqrt{2}} \left(\partial_{\tilde x} +\frac{i}{\tilde G}\partial_{\tilde z}\right) .
\end{eqnarray}

The nonvanishing spin coefficients are
\begin{eqnarray}
\fl\quad \mu &=&\frac{A^2 \tilde F}{2\rho},\quad  
\tau =\frac{A}{\sqrt{2}}\tilde G^{1/2}=-\pi, 
\quad \rho=A(\tilde x + \tilde y), \quad \beta =\frac{\rho}{4\sqrt{2}} \frac{\tilde G'}{\tilde G^{1/2}},
\nonumber \\
\fl\quad \alpha &=&\frac{A\tilde G^{-1/2}}{4\sqrt{2}}[\tilde G -\tilde y \tilde G' +3-\tilde x^2],\quad 
\gamma =\frac{A}{4 (\tilde x+ \tilde y)}[\tilde F +\tilde x \dot {\tilde F} +3-\tilde y^2],
\end{eqnarray}
while the only surviving Weyl scalar is $\psi_2=-mA^3(\tilde x+\tilde y)^3$; here a prime and a dot denote differentiation with respect to $\tilde x$  and $\tilde y$, respectively.
Following the approach of Prestidge \cite{prest}, rescaling the unknown $\psi_s$ of the master equation (for the various $\psi_s$ satisfying the master equation, see e.g. \cite{KNTN}) we find that
\begin{equation}
\label{psiteuk}
\psi_s=(\tilde x+\tilde y)^{(2s+1)}e^{-i\omega t}e^{ik_3\tilde z}X_s(\tilde x) Y_s(\tilde y) 
\end{equation}
gives separated equations for $X_s$ and $Y_s$:
\begin{eqnarray}
\label{eqxy}
\fl
&&  X_s''+\frac{\tilde G '}{\tilde G}  X_s'+
[\frac{-4S-s^2+2p  \tilde x (s^2-4)}{4\tilde G} \nonumber \\
\fl&&  \qquad -
\frac{(-24 p  k_3+ s) s\tilde x^2+2s(9p  s -4k_3)\tilde x  +3s^2+4k_3^2}    
{4\tilde G^2}]X_s=0, \nonumber \\
\fl && \ddot Y_s +\frac{\dot {\tilde F} (s+1)}{\tilde F}  \dot Y_s+
[\frac{S+s(s+1)-2p  \tilde y (s+1)(2s+1)}{\tilde F}+
\frac{\omega (\omega-is\dot{ \tilde F})}
{\tilde F^2}] Y_s=0, 
\end{eqnarray}
where  $S$ is a separation constant.
Because of the symmetry of the metric under the exchange of $\tilde x$  and $\tilde y$, one would expect a similar property to hold for these two equations. It can be shown that this is exactly the case (modulo further replacement of $\tilde y\to -\tilde x$, $\omega\to ik_3$, $s\to -s$) when one uses  the following  rescaling for $X_s(\tilde x)$ and $Y_s(\tilde y)$:
$X_s(\tilde x) \to  X_s(\tilde x)/\tilde G^{1/2}, \quad Y_s(\tilde y) \to  Y_s(\tilde y)/\tilde F^{(s+1)/2}$.
Thus, without any loss of generality one can consider the equation for $X_s$ only. 
This equation, in turn, cannot be solved exactly, unless $p =0$.
In this limit, with $\tilde x=\cos\theta$, one gets
\begin{equation}
\frac{\rmd^2 X_s}{\rmd \theta^2}+\cot \theta  \frac{\rmd X_s}{\rmd \theta}- \left[S+\frac{s^2-2k_3s\cos \theta +k_3^2}{\sin^2\theta}\right]X_s=0,
\end{equation}
so that with $S=-l(l+1)$ and $\tilde z=\phi$, it is easy to show that $X_s(\tilde x)e^{ik_3\tilde z}$ reduces to the standard spin-weighted spherical harmonics. 

Let us consider then the equation for $X_s$ in (\ref{eqxy}), where we set $\tilde x=\cos\theta$ and use the rescaling 
$X_s(\theta)=\sin \theta\mathcal{T}_s(\theta)/\tilde G{}^{1/2}$.
The equation for $\mathcal{T}_s$ is then
\begin{equation}
\label{eq:A8}
\frac{\rmd^2 \mathcal{T}_s}{\rmd \theta^2}+\cot \theta  \frac{\rmd \mathcal{T}_s}{\rmd \theta}- \mathcal{V}\mathcal{T}_s=0,
\end{equation}
where  $\mathcal{V}$ is given by
\begin{equation}
\mathcal{V}=\frac{1}{(1-2p \cos \theta \cot^2 \theta)^2}\left[
p^2 \mathcal{V}_{(2)}(\theta)  +p  \mathcal{V}_{(1)}(\theta) +\mathcal{V}_{(0)}(\theta)
\right]
\end{equation}
and the coefficients
\begin{eqnarray}
\fl\quad \mathcal{V}_{(2)}(\theta)&=& (1-s^2)\cos^2\theta -(1+s^2)\cot^2 \theta +4\cot^6 \theta,\nonumber \\
\fl\quad \mathcal{V}_{(1)}(\theta) &=&2 \cos \theta \left[ 2s^2 (1+\cot^2\theta)-S\cot^2\theta-2(1+\cot^2\theta)^2\right]-6k_3s\cot^2\theta ,\nonumber \\
\fl\quad \mathcal{V}_{(0)}(\theta)&=&  S+\frac{s^2-2k_3 s \cos \theta +k_3^2}{\sin^2\theta} ,
\end{eqnarray}
do not depend on $p$. We recall that in the case under 
consideration here $p < 1/ (3\sqrt{3})$. For $p \ll 1$, it is straightforward to develop a perturbation  series solution to equation (\ref{eq:A8}) in powers of $p$.
In this way, 
terms of the form $ps = msA$ and higher order appear in $\mathcal{V}$, but a
{\it direct} spin-acceleration coupling term $sA$ 
that would be independent of mass $m$ does not exist in $X_s$ and hence $\psi_s$;
therefore, we may conclude that this coupling does not exist. 
To see this in a more transparent way we will consider in the next section a linearization of the C metric.

\section{Linearized C metric}

In metric (\ref{cmetu}) let us consider the coordinate transformation $\{u,r,\theta, \phi\}\rightarrow \{T,X,Y,Z\}$, 
where 
\begin{eqnarray}
\label{approxuXYZ}
&&T=-u- \left[r+2m \ln \left(\frac{r}{2m}-1\right)\right]-Ar^2\cos \theta, \nonumber \\
&&X=r \sin \theta \cos \phi, \quad
Y=r \sin \theta \sin \phi, \quad
Z=r\cos \theta +\frac12 Ar^2.
\end{eqnarray}
The transformed metric becomes
\begin{equation}
\label{newmet}
\rmd s^2= (1-\frac{2m}{R}-2AZ)\rmd T^2 -\frac{2m}{R^3} (X\rmd X+Y\rmd Y+Z\rmd Z )^2 -\rmd X^2 -\rmd Y^2 -\rmd Z^2 ,
\end{equation}
where $R=\sqrt{X^2+Y^2+Z^2}$ and we have neglected $m^2$, $mA$, $A^2$ and higher - order terms.
Next, introduce polar coordinates $\Theta$ and $\Phi$ such that
\begin{eqnarray}
X=R\sin \Theta \cos \Phi,\quad Y=R \sin \Theta \sin \Phi,\quad
Z=R\cos \Theta .
\end{eqnarray}
With respect to these, metric (\ref{newmet}) becomes
\begin{equation}
\label{linearizzata}
\rmd s^2= \left(1-\frac{2m}{R}-2AR\cos \Theta \right)\rmd T^2 -\left(1+\frac{2m}{R}\right) \rmd R^2 -R^2 (\rmd \Theta^2 +\sin^2 \Theta \rmd \Phi^2) .
\end{equation}
Finally, introducing the isotropic radial coordinate $\rho$,
\begin{equation}
R=\left(1+\frac{m}{2\rho}\right)^2\rho=\rho +m +\frac{m^2}{4\rho}\simeq \rho+m\, ,
\end{equation}
we get the linear metric in standard form
\begin{equation}
\label{iso}
\rmd s^2= \left(1-\frac{2m}{\rho}-2A\hat Z\right)\rmd T^2 -\left(1+\frac{2m}{\rho}\right) (\rmd \hat X{}^2 +\rmd \hat Y{}^2+\rmd \hat Z{}^2), 
\end{equation}
where
\begin{equation}
 \hat X= \rho \sin \Theta \cos \Phi , \quad \hat Y= \rho \sin \Theta \sin \Phi, \quad \hat Z= \rho \cos \Theta .
\end{equation}

\subsection{Gravitational Stark effect}

Consider the massless scalar field equation
$\nabla^\mu \partial_\mu \chi =0$
on the background spacetime given by the metric (\ref{iso}). 
To first order in $m$ and $A$, $\chi$ can be separated by introducing parabolic coordinates  in analogy with the Stark effect, which is
the shift in the energy levels of an atom in an external electric field corresponding to the eigenvalues of a Schr\"odinger equation with
a Coulomb potential $-k/r$ plus the potential due to a constant electric field $\mathbf{E}=E \, \hat{\mathbf{z}}$, i.e. $-k/r + eE z$, where  $-e$ is the charge of the electron. 
In this gravitoelectromagnetic counterpart of the Stark effect, we set
\begin{equation}
\hat X = \sqrt{\xi \eta} \cos \psi, \quad \hat Y = \sqrt{\xi \eta} \sin \psi, \quad \hat Z =\frac12 (\xi-\eta), 
\end{equation}
and assume that
\begin{equation}
\label{Psi}
\chi(T,\xi,\eta,\psi)= e^{-i \omega T}\,  e^{i \nu \psi} U(\xi) V(\eta),
\end{equation}
where $\xi\ge 0$, $\eta\ge 0$, $\psi$ takes values from $0$ to $2\pi$, $\omega$ is a constant and $\nu$ is an integer.

It follows from the scalar wave equation that
\begin{eqnarray}
\label{eqUV}
&& U_{\xi\xi}+\frac{1}{\xi}\left( 1-\frac12 A\xi \right)U_\xi+\left[\frac{\omega^2}{4}(1+\xi A) + \frac{1}{\xi}(m\omega^2 -C)-\frac{\nu^2}{4\xi^2}\right]U=0 , \nonumber \\
&& V_{\eta\eta}+\frac{1}{\eta}\left( 1+\frac12 A\eta \right)V_\eta+\left[\frac{\omega^2}{4}(1-\eta A) + \frac{1}{\eta}(m\omega^2 +C)-\frac{\nu^2}{4\eta^2}\right]V=0,
\end{eqnarray}
where $C$ is the separation constant and $U_\xi=\rmd U /\rmd \xi$, etc.
Note that the second equation for  $V(\eta)$ can be obtained from the first one for $U(\xi)$ by replacing $A\to -A$ and $C\to -C$.
Introducing a new constant $\beta$ by
\begin{equation}
C=\frac12\left(\beta -\frac{A}{2}\right)
\end{equation}
and rescaling $U$ and $V$,
\begin{equation}
U(\xi) =\left(1+\frac{A\xi}{4}\right)a(\xi) , \qquad V(\eta)=\left(1-\frac{A\eta}{4}\right)b(\eta) ,
\end{equation}
eqs. (\ref{eqUV}) become
\begin{eqnarray}
\label{eqab}
&& \frac{d}{d\xi} \left(\xi \frac{da}{d\xi}\right)+ \left[\frac{\omega^2 \xi}{4}-\frac{\nu^2}{4\xi}+\frac{A\omega^2}{4}\xi^2 +\left(m\omega^2+\frac{\beta}{2}\right)\right]a=0,\nonumber \\
&& \frac{d}{d\eta} \left(\eta \frac{db}{d\eta}\right)+ \left[\frac{\omega^2 \eta}{4}-\frac{\nu^2}{4\eta}-\frac{A\omega^2}{4}\eta^2 +\left(m\omega^2-\frac{\beta}{2}\right)\right]b=0.
\end{eqnarray}

These equations can be put in exact correspondence with the  Schr\"odinger equation for the hydrogen atom in a constant electric field that results in the Stark effect \cite{LLMQ}. For details see \cite{bcmprd}.

Let us note here again  the close formal correspondence between the  quantum theory of the Stark effect in hydrogen and the theory of a classical massless scalar field on the linearized C-metric background. 
Extension of this result to massless fields with nonzero spin present difficulties, as we have already seen in section IV.

Finally, for many laboratory applications, the potential associated with the gravitational Stark effect can be written as $m/\rho+A\rho \cos \Theta$ with
$\rho=\rho_\oplus+\zeta$, where $\rho_\oplus$ is the average radius of the Earth and $\zeta$ is the local vertical coordinate in the laboratory. Using the local acceleration  of gravity, $g=m/\rho_\oplus^2$, the effective {\it Newtonian} gravitational potential is then $-m/\rho_\oplus +g\zeta -A(\rho_\oplus+\zeta)\cos \Theta$; some of the applications of this potential are discussed in the next subsection.

\subsection{Acceleration-induced phase shift}

From the gravitational Stark effect we have learned that  wave phenomena in the exterior spacetime represented by (\ref{iso}) are affected by the acceleration $A$. Consider then wave fields in a laboratory fixed on the Earth
assumed to undergo a small uniform nongravitational acceleration (e.g. solar radiation pressure or Mathisson - Papapetrou coupling of the curvature of the solar gravitational field with the angular momentum of the Earth).
Estimates suggest that such accelerations are very small and at a level below $\sim 10^{-10}$ cm/s$^2$. 
In this sense, the total field of the Earth (nonrotating, spherical and endowed with a very small acceleration) is taken into account by the linearized C metric and we expect that
the Earth's acceleration will introduce a very small shift in the phase of a wave propagating in the gravitational field of the Earth. Consider, for instance, the gravitationally induced quantum interference of neutrons as in the COW experiment \cite{colella, RauchandWerner}. Let us imagine for the sake of simplicity that the $\hat Z$- axis of the system $\{ T, \hat X, \hat Y, \hat Z\}$ of metric (\ref{iso}) 
makes an angle $\Theta$ with the vertical direction in our local laboratory and so an otherwise free particle in the laboratory is subject to the effective Newtonian gravitational acceleration $g-A\cos \Theta$.
The corresponding neutron phase shift in the COW experiment would then be given by
\begin{equation}
\label{cownew}
\Delta \varphi =(g - A\cos\Theta) \frac{\mathcal{A}\omega}{v}\sin\alpha ,
\end{equation}
where $\omega$ is the de Broglie frequency of the neutron,  $\mathcal{A}$ is the area of the interferometer, $\alpha $ is the inclination angle of the interferometer plane with respect to the horizontal plane in the laboratory and $v$ is the neutron speed. When $A = 0$ or $\Theta=\pi/2$, this formula reduces to the standard formula of the COW experiment \cite{RauchandWerner}. A complete discussion of the neutron phase shift for nonzero $A$ is beyond the scope of this work.

\subsection{Pioneer anomaly}
Imagine an inertial reference frame and a star of mass $m$ such that its center of mass accelerates with a constant acceleration $\mathbf{A}=A \hat{\mathbf{z}}$ with $A>0$.
Thus the motion of a planet or a satellite about the star in terms of a noninertial coordinate system $\{t,x,y,z\}$ in which the star is at rest with its center of mass at the origin of the spatial coordinates is given to lowest order by
\begin{equation}
\frac{\rmd^2 \mathbf{r}}{\rmd t^2}+ \frac{m \mathbf{r}}{r^3}=-\mathbf{A}
\end{equation}
in accordance with Newtonian physics. Within the context of general relativity, the equation of motion of the test planet or satellite is given by the geodesic equation in the vacuum C metric.

Let us now apply these ideas to the anomalous acceleration of Pioneer 10 and Pioneer 11 \cite{anderson1, anderson2, andmas}, 
launched over thirty years ago to explore the outer solar system.
The analysis of Doppler tracking data from Pioneer 10/11 spacecraft (moving away from the solar system in almost opposite directions) is consistent with the existence of a small anomalous acceleration of about $10^{-7}\,$ cm/s${}^2$ toward the Sun. 

Let $\hat{\mathbf{P}}$ and $\hat{\mathbf{P}}'$ be unit vectors that indicate the radial directions of motion of Pioneer 10 and Pioneer 11 with respect to the Sun, respectively. Suppose that the smaller angle between these directions is given by $\pi - 2 \beta$, where $\beta \simeq 7^\circ$. Then, $\mathbf{A}$ can be expressed as
\begin{equation}
\label{eq:56}
        \mathbf{A} = \frac{A_0}{2 \sin \beta }(\hat{\mathbf{P}}+\hat {\mathbf{P}}'), 
\end{equation}
where $A_0 \simeq 10^{-6}$ cm/s${}^2$ is the magnitude of the vector $\mathbf{A}$ and is such that, with $\sin 7^\circ \simeq 0.12$, $A_0 \sin \beta$ is the magnitude of the anomalous acceleration.
It follows that $\mathbf{A}\cdot \hat{\mathbf{P}} = \mathbf{A}\cdot \hat{\mathbf{P}}' = A_0 \sin \beta$.
It is therefore possible to find a vector $-\mathbf{A}$ that generates the Pioneer anomaly; however, the problem is then shifted to explaining the origin of such an acceleration of the center of mass of the Sun.

One possibility could be recoil acceleration due to the anisotropic emission of solar radiation.
But estimates for this effect give $A<  10^{-10}$ cm/s${}^2$, so that it does not appear possible to explain the Pioneer anomaly in this way.


\begin{thebibliography}{00}

\bibitem{srcoupl}
B. Mashhoon, Phys. Rev. Lett. {\bf 61}, 2639 (1988).  

\bibitem{LC}
T. Levi-Civita, Rend. Accad. Naz. Lincei {\bf 27}, 343 (1918).

\bibitem{newtam}
E. Newman  and L. Tamburino,  J.\, Math.\, Phys. {\bf 2}, 667 (1961).

\bibitem{robtra}
I. Robinson and A. Trautman, Proc. Roy. Soc.\/ (London) {\bf A265}, 463 (1962).

\bibitem{ehlkun}
J. Ehlers and W. Kundt,
in {\it Gravitation: An Introduction to Current Research}, 
ed L. Witten, Wiley,  New York (1962).

\bibitem{kin69}
W. Kinnersley,  Phys.\, Rev.\/ {\bf 186}, 1335 (1969).

\bibitem{kinwal}
W. Kinnersley  and M. Walker,  Phys.\, Rev. D\/ {\bf 2}, 1359 (1970).

\bibitem{bon83}
W.B. Bonnor,  Gen.\ Rel.\ Grav.\/ {\bf 15}, 535 (1983).

\bibitem{ES}
H. Stephani, D. Kramer, M.A.H.  MacCallum, C. Hoenselaers  and E. Herlt,   
{\it Exact Solutions of Einstein's Theory}, 
Cambridge Univ.\ Press, Cambridge, second edition (2003).

\bibitem{Farh}
H. Farhoosh  and L. Zimmerman, Phys. Rev. D {\bf 21}, 317 (1980).

\bibitem{pavda}
V. Pravda  and A. Pravdov\'a, Czech. J. Phys. {\bf 50}, 333 (2000).

\bibitem{podol}
J. Podolsk\'y  and J.B. Griffiths, Gen.\ Rel.\ Grav.\/ {\bf 33}, 59 (2001).

\bibitem{bcgj}
D. Bini, C.  Cherubini, A. Geralico, R.T. Jantzen, in preparation (2004).

\bibitem{Guven} 
R. G\"uven,   
Phys.\ Rev.\ D {\bf  22}, 2327 (1980).

\bibitem{teuk1}
S.A. Teukolsky,  
Phys.\ Rev.\ Lett.   {\bf29}, 1114 (1973).

\bibitem{teuk2}
S.A. Teukolsky,  
Astrophys.\ J.   {\bf 185}, 635  (1973).

\bibitem{Detweiler} 
S.L. Detweiler  and J.R. Ipser,  
Astrophys. J.\   {\bf 185}, 675  (1973).

\bibitem{Wainwright} 
J. Wainwright  
J.\ Math.\ Phys.\   {\bf 12}, 828 (1971).

\bibitem{RARITA2} 
G.F.T. Del Castillo  and  G. Silva-Ortigoza,   
Phys. Rev.\ D  {\bf   42}, 4082 (1990).

\bibitem{Finley2}
A.L. Dudley  and J.D. Finley  III, 
Phys. Rev. Lett.\ {\bf
38}, 1505 (1977); Errata: Phys. Rev. Lett.\/ {\bf
39}, 367 (1977).

\bibitem{prest}
T. Prestidge, Phys.\, Rev.\/ D {\bf 58}, 124022 (1998).

\bibitem{bcmprd}
D. Bini, C. Cherubini and B. Mashhoon, 
Phys.\, Rev.\ D, {\bf 70}, 044020 (2004).

\bibitem{bcmcqg}
D. Bini, C. Cherubini and B. Mashhoon, 
Class. Quantum Grav. {\bf 21}, 3893 (2004).

\bibitem{KNTN}
D. Bini, C. Cherubini, R.T. Jantzen and B. Mashhoon, 
Phys.\, Rev.\ D {\bf 67}, 084013 (2003).

\bibitem{LLMQ}
L.D. Landau and E.M. Lifshitz,  {\it Quantum Mechanics}, 
Pergamon Press, Oxford (1965).

\bibitem{colella}
R. Colella, A.W. Overhauser  and  S.A. Werner,  
Phys.\ Rev.\ Lett.\/ {\bf34}, 1472 (1975).

\bibitem{RauchandWerner} 
H. Rauch  and  S.A. Werner, 
{\it Neutron Interferometry}, Clarendon Press, Oxford (2000).

\bibitem{anderson1}
J.D. Anderson  et al., Phys. Rev. Lett.\/ {\bf 81}, 2858 (1998).

\bibitem{anderson2}
J.D. Anderson  et al., Phys. Rev.\ D {\bf 65}, 082004 (2002).

\bibitem{andmas}
J.D. Anderson  and B. Mashhoon, Phys. Lett. \ A {\bf 315}, 199 (2003).



\end{thebibliography}
\end{document}